\documentclass[conference]{IEEEtran}
\IEEEoverridecommandlockouts

\usepackage{cite}
\usepackage{amsmath,amssymb,amsfonts}
\usepackage{algorithmic}
\usepackage{graphicx}
\usepackage{textcomp}
\usepackage{xcolor}
\def\BibTeX{{\rm B\kern-.05em{\sc i\kern-.025em b}\kern-.08em
    T\kern-.1667em\lower.7ex\hbox{E}\kern-.125emX}}
\begin{document}

\newcommand*\CX{\color{red}CX:}

\title{Soil Analysis with Machine-Learning-Based Processing of Air-Coupled Stepped-Frequency GPR Field Measurements: Preliminary Study\\
}

\author{
    Chunlei Xu\IEEEauthorrefmark{1}, 
    Michael Pregesbauer\IEEEauthorrefmark{2}, 
    Naga Sravani Chilukuri\IEEEauthorrefmark{1}, 
    Daniel Windhager\IEEEauthorrefmark{1}, \\
    Mahsa Yousefi\IEEEauthorrefmark{2},
    Pedro Julian\IEEEauthorrefmark{3},
    Lothar Ratschbacher\IEEEauthorrefmark{1}\\
    \IEEEauthorblockA{\IEEEauthorrefmark{1} \textit{Intelligent Wireless Systems Division, Silicon Austria Labs GmbH}, Linz, Austria }
    \IEEEauthorblockA{\IEEEauthorrefmark{2} \textit{Geoprospectors GmbH}, Traiskirchen, Austria}
    \IEEEauthorblockA{\IEEEauthorrefmark{3} \textit{Electrical Engineering Research Institute, Universidad Nacional del Sur}, Bahia Blanca, Argentina }
    \IEEEauthorblockA{Email: chunlei.xu@silicon-austria.com, michael@pregesbauer.com, daniel.windhager@silicon-austria.com, \\ pjulian@uns.edu.ar, lothar.ratschbacher@silicon-austria.com}
}

\maketitle

\begin{abstract}

Ground Penetrating Radar (GPR) has been widely studied as a tool for extracting soil parameters relevant to agriculture and horticulture. When combined with Machine Learning (ML) methods, air-coupled Stepped Frequency Continuous Wave Ground Penetrating Radar (SFCW GPR) measurements could offer a cost-effective way to obtain depth-resolved soil data. As a first step of our study in this direction, we conducted an extensive field survey using a tractor-mounted air-coupled SFCW GPR instrument. Leveraging ML-based data processing, we evaluate the GPR instrument's ability by predicting the apparent electrical conductivity (ECaR) measured by a co-recorded Electromagnetic Induction (EMI) instrument. The large-scale field measurement campaign with 3472 co-registered and geo-located GPR and EMI data samples distributed over approximately 6600 square meters was performed on a golf course. This terrain offers high surface homogeneity but also presents the challenge of subtle soil parameter variations. Based on the results, we discuss challenges in this multi-sensor regression setting and propose the use of the nugget-to-sill ratio as a performance metric for evaluating ML models in agricultural field survey applications.

\end{abstract}

\begin{IEEEkeywords}
Stepped frequency ground penetrating radar, Electromagnetic induction, Machine learning, Soil analysis. 
\end{IEEEkeywords}

\section{Introduction}
Accurate soil information is crucial for effective land management, from resource-efficient agri- and horticulture to hydrological hazard mitigation~\cite{Zhuo2019}. Depending on the spatio-temporal scale and accuracy requirements for soil information, different measurement methods are in use today. At the local level, (networks of) sensors provide direct, quasi instantaneous measurements of soil parameters, including depth-resolving measurements~\cite{Francia2022}, whereas at large
scales, satellite-based systems with longer revisit times have been employed~\cite{Liang2021}. As the focus of this paper, tractor-mounted instruments offer a resource-efficient solution for real-time soil data collection, enabling to directly determine process parameters in time for subsequent (agri- and horticultural) equipment. The two most established non-invasive soil sensing techniques are Electromagnetic Induction (EMI) and Ground Penetrating Radar (GPR). Extensive research has explored how to relate both instrument readouts with soil parameters, including soil layer thickness, soil density and soil water content (SWC)~\cite{Huisman2003, Tran2012, Klotzsche2018, Pathirana2023, Visconti2021, Lombardi2022}.

EMI instruments measure apparent electrical conductivity (ECaR) in the surrounding soil using low-frequency (VLF) electromagnetic waves. The probing frequency and the specific coil configuration (spacing, orientation, and height) determine the instrument sensitivity to subsurface layers at various depths for different soil volumes~\cite{hess_EMI2022, Schmack_2022}. GPR, on the other hand, provides depth-resolved soil sensing by transmitting pulsed or frequency-stepped/modulated (VHF-UHF) radio frequency waves to the ground~\cite{Diamanti2017, Pawel2019, Lombardi2019}. The selection of TX and RX antenna configurations depends on the target application, including air-coupled vs. ground-coupled setups, antenna orientation and monostatic, fixed-offset or variable-offset arrangements. The choice of radar center frequency involves a trade-off: high frequencies improve spatial resolution but reduce penetration depth and increase sensitivity to surface roughness. Stepped frequency continuous wave (SFCW) GPR, recording over a series of dwell times at discrete frequency steps, offers flexible control adjustable to the desired resolution and ground conditions, enabling dynamic penetration depths, a wide bandwidth and high signal-to-noise ratios (SNRs)~\cite{Oyan2012, Eide2014, Koganti2020}. 

However, both EMI and GPR instruments require site-specific calibration, since first principle approaches without free parameters tend to yield precise results only for well-defined laboratory settings. In the field, challenges arise due to soil heterogeneity, vegetation, surface morphology, and interference from nearby machinery. Recently, machine learning (ML) has shown potential in enhancing GPR analysis and soil parameter estimation. For example, deep neural networks such as GPRNet~\cite{Leong2021} have been used for direct velocity inversion, and have in further developments been applied to depth-resolved SWC profiling~\cite{Li2023} in the agricultural context. ML-based methods have also optimized classical signal processing features for soil moisture estimation in a field trial under controlled moistening conditions with multiple geophysical instruments, including low-frequency GPR~\cite{Terry2023}. For more precise depth-resolved soil parameter estimation, high-frequency SFCW radar combined with ML-based processing has shown promising results based on data from multiple field locations~\cite{Filardi2023}.

Although direct field sampling remains the gold standard for soil parameter characterization, it is often impractical for the large dataset acquisition needed to develop and validate data-driven methods. In this paper, we thus adopt the data acquisition approach from \textit{Jonard et al.}~\cite{Jonard2013} to record a large field area with both EMI and GPR instruments simultaneously.
We aim to assess the feasibility of soil sensing with an air-coupled SFCW GPR instrument and end-to-end  ML methods, using co-measured EMI data from the large-scale campaign\footnote[1]{The code of this paper is available at: https://opensource.silicon-austria.com/xuc/soil-analysis-machine-learning-stepped-frequency-gpr.} as a proxy for relevant soil parameters.

\section{Materials and Methods}

\subsection{Sensor Specification and Field Campaign}

\begin{figure}[!htb] 
	\centering
		\includegraphics[width=0.45\textwidth, height=4.2cm]{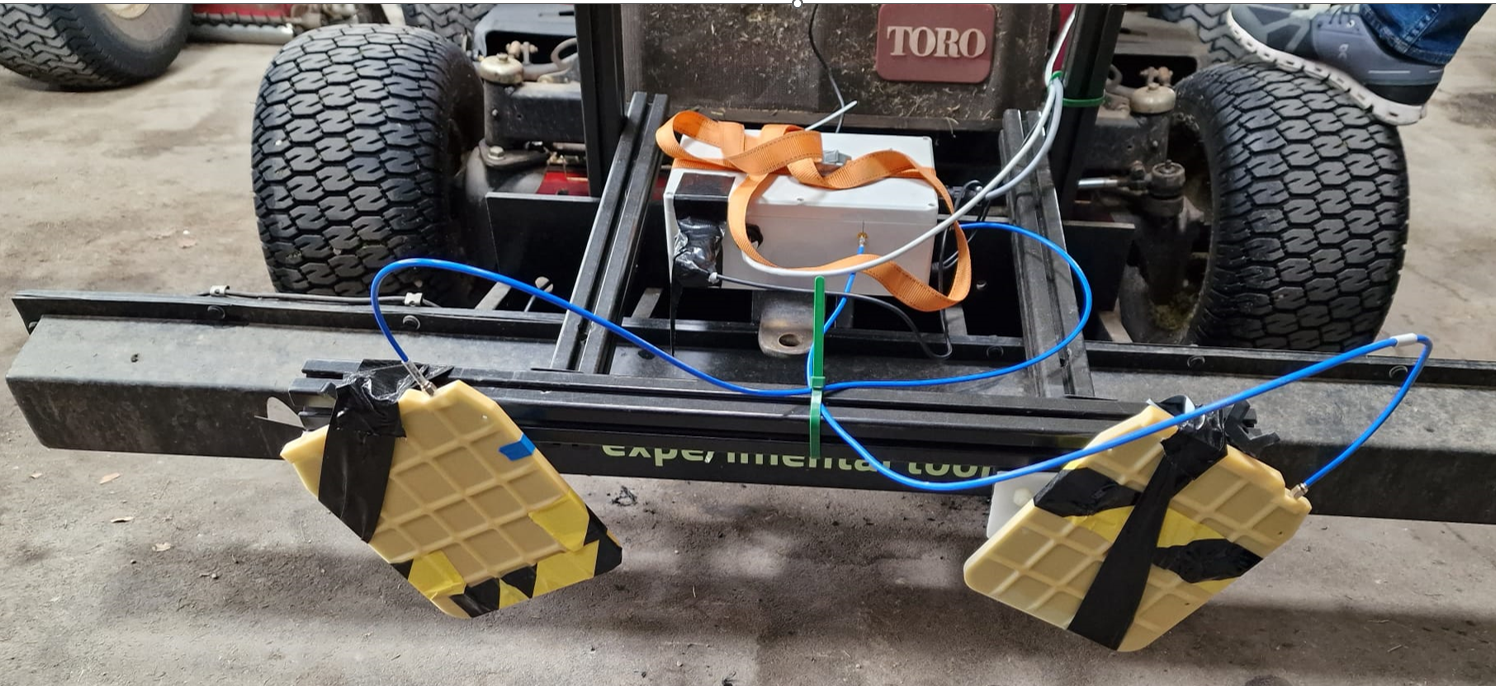}
	\caption{Experimental setup of the air-coupled SFCW GPR with a fixed offset and Vivaldi antennas (in yellow color) mounted directly behind the transversal EMI instrument bar on a \textit{Toro Reelmaster 5510} tractor.}    
	\label{FIG:setup}
\end{figure}

For the field measurement, a tractor (\textit{Toro Reelmaster 5510}) was equipped with a \textit{Topsoil Mapper} EMI instrument by \textit{Geoprospectors} and a newly developed SFCW GPR radar (see Fig.~\ref{FIG:setup}). The system was integrated with a Real-Time Kinematic (RTK) corrected GPS (\textit{Stonex s10a}) unit recording geo-location and time stamping information at a rate of 1\,Hz and with a single point accuracy of 2-3\,cm. The EMI instrument operated with a horizontal coil alignment at a frequency of 9\,kHz and was mounted 20~cm above the ground, recording \textit{raw} ECaR values at a sampling rate of 5\,kHz. Here, \textit{raw} indicates that measurements included a baseline offset caused by the presence of  conductive material from the mower, which was not removed by calibration in this study.

\subsubsection{SFCW GPR}
A single channel SFCW radar prototyped by \textit{Geoprospectors} is built as a bistatic, air-coupled system with fixed-offset Vivaldi antennas for both transmission and reception (detailed specifications in Table~\ref{TABLE:SFCW}).

\subsubsection{Study Site and Data Source}
\label{sec:studysite}
\begin{figure*}[h]
	\centering
		\includegraphics[width=0.95\textwidth, height=5.8cm]{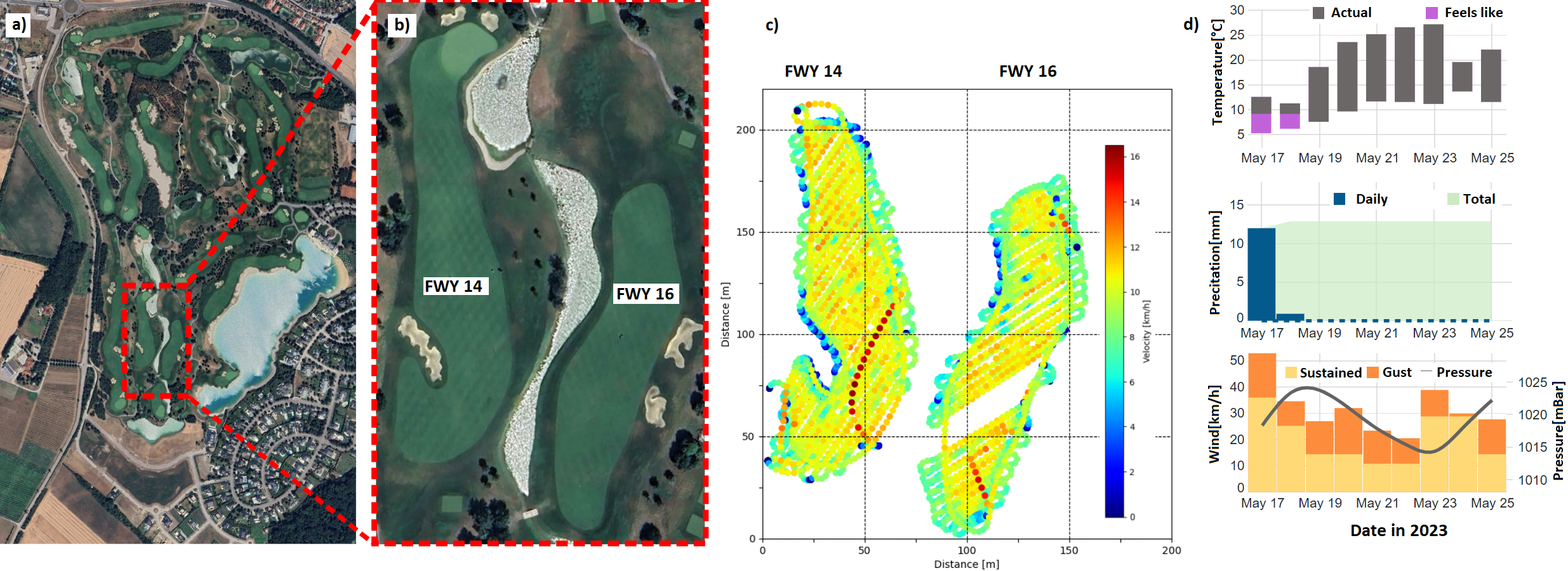}
	\caption{Information of the field campaign. a) Overview satellite map and b) detailed satellite imagery of the Fontana Golf Club, Austria. c) Velocity map of the tractor on FWY14 (left) and FWY16 (right). d) Weather conditions in the week leading up to the date that measurements were taken.}    
	\label{FIG:studysite}
\end{figure*}

The field campaign took place on fairway 14 (FWY14) and fairway 16 (FWY16) of the Fontana Golf Club, Austria (47°58'29"N, 16°18'25"E, $\sim$220m elevation) on 25.05.2023,  following several days of dry, warm and windy weather. Data was collected along parallel lanes of fairways ($\sim$1.5\,m spacing) with the tractor driving at speeds of up to 16~km/h (see Fig.~\ref{FIG:studysite}). The EMI, SFCW GPR and GPS data were co-registered and re-sampled to match the SFCW GPR sampling rate.

\begin{table}[h]
\caption{SFCW GPR Setup Parameters}
\label{TABLE:SFCW}
\begin{center}
\begin{tabular}{|l|l|}
\hline
\textbf{Parameter} & \textbf{Value}  \\
\hline
Frequency & 1.3--2.9 GHz \\
\hline
Total sweep time & 70 ms\\
\hline
Number of steps & 400 \\
\hline
Max. radar location sampling rate & 10 Hz \\
\hline
Antenna separation & 60 cm at feed points \\
\hline
Ground clearance & 15 cm \\
\hline
Angle to vertical & $23^{\circ}$ for both antennas \\
\hline
Antenna Gain & $\sim$7 dBI constant over bandwidth \\
\hline
\end{tabular}
\end{center}
\end{table}

\subsection{Machine Learning}
To evaluate the effectiveness of high-resolution air-coupled SFCW GPR in soil sensing, the EMI instrument can serve as a proxy for this purpose. The goal is to predict ECaR values of EMI from SFCW GPR data using data-driven ML models.

\subsubsection{Data Preprocessing}
\label{sec:dataprocess}
The field dataset, including radar and ECaR measurements, is spatial and temporal filtered to reduce instrument response variations caused by tractor turns and velocity changes. Data samples associated to turning paths, non-parallel sampling paths at the beginning and end of recorded sequences, and extreme velocity values are removed. The dataset is then prepared for the training and testing of ML models, where radar readings at the 400 frequency steps serve as inputs, and measured ECaR values are used as targets at each geolocation. Outliers in the upper and lower 0.5\% of ECaR values are excluded, resulting in a total number of $3472$ samples for both FWY14 and FWY16 fairways. Each sample has radar values, a target ECaR value, geographic coordinates, and an estimated tractor speed. Radar data is normalized by subtracting the mean value over all samples for each frequency step, following a similar approach as~\cite{Filardi2023}. Furthermore, a transformation that yields zero mean and standard deviation of one (for the target variables on the training dataset) is applied to all target variables with the purpose to improve learning during the regression task. 

\subsubsection{Regression Models and Performance Metrics}

Regression models—Linear Regression (Linear), Random Forest Regression (RFR) and $k$-nearest-neighbor-based regression (KNR)—are employed to predict ECaR values from SFCW GPR data, with Mean Squared Error (MSE) as the loss function. Model performance is evaluated using MSE, Mean Absolute Error (MAE), Pearson correlation coefficient ($r$) and Nugget-to-Sill Ratio (NSR) derived from variograms~\cite{varigram_cressie}. Hyperparameters for RFR and KNR are optimized using grid search in a (repeated) nested cross-validation (CV) setup~\cite{Bates2023, ryai_nest_cv, Varma2006BiasIE} to select the best architectures based on MSE. Models are then trained and evaluated using repeated CV to ensure robust error estimation. Performance metrics and their one-sigma confidence intervals are calculated based
on 50 evaluations that arise from the ten repetitions of the 5-fold cross validation.

\section{Results}
\label{sec:results}
\begin{figure*}[htb!]
    \centering
    \raggedleft
	\includegraphics[width=0.95\textwidth, height=4.8cm]{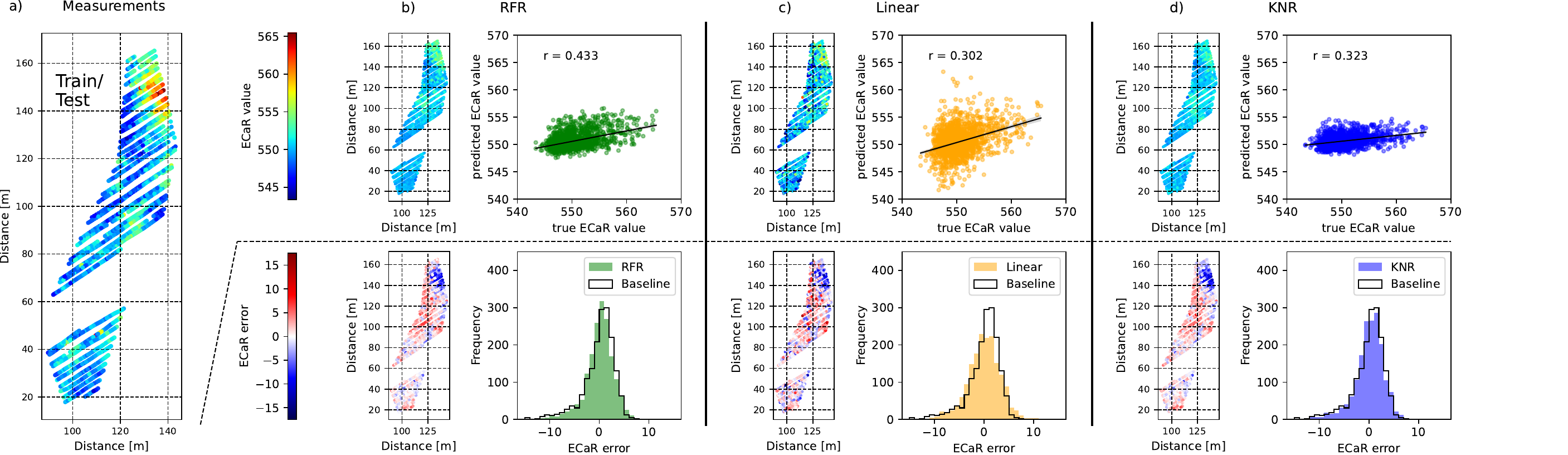}
	\caption{Results of predicted ECaR values from SFCW GPR data in Scenario 1. A geo-randomized five-fold cross-validation is applied to measurements of FWY16 with measured ECaR values plotted in heatmap a). Results of RFR, Linear regression and KNR models are shown in b)-d). The top row presents model predictions in heatmaps and scatter plots between measured ECaR values and predictions (with linear fit and Pearson correlation coefficient $r$), while the bottom row displays model prediction errors in heatmaps and histograms. The histograms of baseline prediction errors represent from a uniform prediction using the mean of ECaR values of a training set.} 
	\label{FIG:GTvsModel16}
\end{figure*}

\begin{figure*}[htb!]
	\centering
    \raggedleft
		\includegraphics[scale=.33]{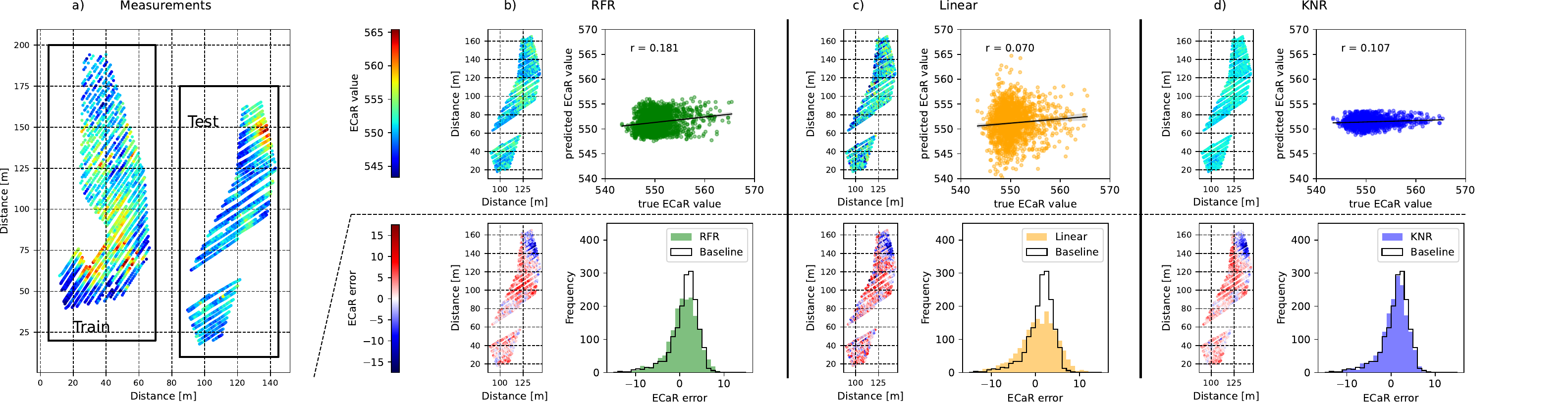}
	\caption{Results of predicted ECaR values from SFCW GPR data in Scenario 2. Models are trained on FWY14 and tested on FWY16 with measured ECaR values plotted in heatmap a). Results for RFR b), Linear regression c), and KNR d) follow the description in Fig.~\ref{FIG:GTvsModel16}.} 
	\label{FIG:GTvsModel1416}
\end{figure*}

Based on data from FWY14 and FWY16 described in Sect.~\ref{sec:studysite}, two scenarios are analyzed in this study.
\begin{itemize}
    \item Scenario 1: Only FWY16 data is used for both training and testing (cf. Fig.~\ref{FIG:GTvsModel16} a)). Hyperparameters of RFR and KNR are optimized on FWY16 data. Models are then trained and tested using repeated 5-fold CV with randomly split data (ignoring location information), resulting in train and test samples being in close spatial proximity to each other.
    \item Scenario 2: FWY14 data is used for training, and FWY16 data for testing (cf. Fig.~\ref{FIG:GTvsModel1416} a)). Hyperparameters of RFR and KNR models are optimized on FWY14 data. Models are then trained using repeated 5-fold CV with four folds from FWY14 and tested on the entire FWY16 dataset, ensuring large spatial separation between train and test samples.
\end{itemize}

In Scenario 1, heatmaps of predicted ECaR values (top row plots of Fig.~\ref{FIG:GTvsModel16} b), c) and d)) exhibit similar geographic patterns to the heatmap of measured ECaR values (Fig.~\ref{FIG:GTvsModel16}a)), but with a lower dynamic range. Scatter plots showing Pearson correlation coefficient values between measured ECaR values and model predictions reveal challenges in estimating extreme ECaR values and highlight variances in prediction errors. Although the Linear model has the highest fitted slope, RFR shows the least variance in prediction error and achieves the highest Pearson correlation coefficient ($r= 0.425$), making it the best-performing model. This is also reflected in the error histograms (bottom row plots of Fig.~\ref{FIG:GTvsModel16}b), c) and d)), where prediction errors of RFR result in a narrower distribution compared to those of the baseline, which simply assumes the average of ECaR values from the training data as a constant prediction for the test data. As summarized in Table~\ref{TAB:ERROR}, RFR performs best overall, while the ranking of models depends on the chosen error metric.

\begin{table*}[h]
  \caption{Summary of performances of RFR, linear regression and KNR models for the studied train/test scenarios.}
  \centering 
  \begin{tabular}{|c| c| c|c |c| c| c| c| c| } 
    \hline

    &\multicolumn{4}{c|}{Train and Test on FWY16} & \multicolumn{4}{c|}{Train on FWY14, Test on FWY16} \\ 
    \hline
     & RFR & Linear & KNR & Baseline & RFR & Linear & KNR & Baseline  \\
    \hline

MAE	&	\textbf{2.11$\pm$0.10}&	2.61$\pm$0.12	&	2.22$\pm$0.10	&	2.31$\pm$0.11	&2.72$\pm$0.06&	3.57$\pm$0.11	&	\textbf{2.60$\pm$0.03}	&	2.70$\pm$0.02	\\
MSE	&	\textbf{8.10$\pm$0.86}	&	12.15$\pm$1.92	&	8.93$\pm$0.88	&	9.94$\pm$1.07	&	11.99$\pm$0.33	&	21.66$\pm$1.38	&	\textbf{10.85$\pm$0.18}	&	11.23$\pm$0.09	\\
r	&	\textbf{0.43$\pm$0.06}	&	0.33$\pm$0.04	&	0.32$\pm$0.05	&	n.a.			        &	\textbf{0.18$\pm$0.02}	&	0.08	$\pm$	0.02	&	0.11$\pm$0.02	&	n.a.			\\
  \hline  
  \end{tabular}
  \label{TAB:ERROR}
\end{table*}

\begin{table*}[h]
  \caption{Variogram parameters extracted from spherical fits to the model predictions and ground truth variograms in Fig.~\ref{FIG:variogram}. the nugget-to-sill ratio (NSR) correlates with performance metrics in Table~\ref{TAB:ERROR}.}
  \centering 
  \begin{tabular}{ |c| c| c| c| c| c| c| c| } 
    \hline
    &\multicolumn{3}{ c |}{Train and Test on FWY16}  &\multicolumn{3}{ c| }{Train on FWY14, Test on FWY16} & Ground Truth FWY16\\ 
    \hline
     & RFR & Linear & KNR  &  RFR & Linear & KNR &  \\
        \hline
    Range    &	17.62$\pm$0.20	&	18.18$\pm$1.09	&	19.09$\pm$0.31	 &	26.10$\pm$1.29		& 17.41$\pm$2.30	&	21.03$\pm$	1.42	&	15.82	\\
    Nugget &	0.67$\pm$0.04	&	5.93$\pm$0.57	&	0.50$\pm$0.02	 &	1.83$\pm$0.26	& 11.35$\pm$1.06	&	0.54$\pm$0.05	&	1.51	\\
    Sill   &	1.78$\pm$0.04	&	7.96$\pm$0.92	&	1.13$\pm$0.02	 &	3.88$\pm$0.47		&  13.00$\pm$1.40	&	0.81$\pm$0.07	&	9.71	\\ 
    NSR     &	\textbf{0.38$\pm$0.02}	&	0.75$\pm$0.02	&	0.44$\pm$0.01	 &	\textbf{0.47$\pm$0.03}	&	0.87$\pm$0.04	&	0.66$\pm$0.02	&	0.16	\\
    \hline
  \end{tabular}
  \label{TAB:variogram}
\end{table*} 

In Scenario 2, spatial separation between training and testing data leads to a general decline in model performance indicated by the reduced slopes in scatter plots (cf. Fig.~\ref{FIG:GTvsModel1416}). RFR still achieves the highest Pearson correlation coefficient for ECaR predictions, but ranks second to KNR in MAE and MSE, with KNR being the only model outperforming the baseline (performance metrics of models compared in Table~\ref{TAB:ERROR}). 

\begin{figure}[!ht]
	\centering
		\includegraphics[width=0.48\textwidth, height=4.3cm]{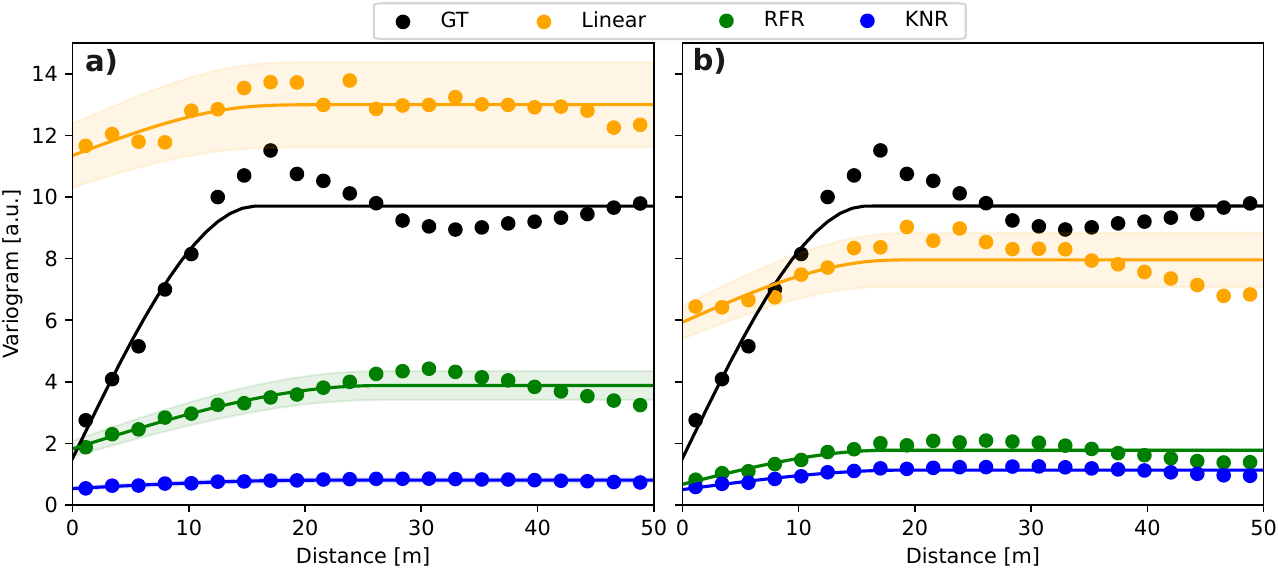}
	\caption{Variograms of Scenario 1 with training and testing on FWY16 a) and Scenario 2 with training on FWY14 and testing on FWY16 b).}   
	\label{FIG:variogram}
\end{figure}

To further evaluate model performance, variograms of predicted ECaR values were calculated and fitted with spherical covariance models (see Fig.~\ref{FIG:variogram} and Table~\ref{TAB:variogram}). Variograms play an essential role in geostatistics, which quantify the spatial variability of soil properties and are used for optimal interpolation of measured data points~\cite{Mzuku_varigram_soil_spatial}. Key parameters—nugget (the (extrapolate) variance between samples at vanishing distances), sill (the variance of samples at (infinitely) large distances), range and NSR—were calculated. Since the nugget value has a lower bound by the intrinsic variance of the measurement/estimation method~\cite{Viscarra1998}, NSR could be served as a meaningful performance metric, which does not require ground truth information for its evaluation. Lower NSR values imply stronger spatial correlation and better model predictions. The measured ECaR values exhibit the lowest NSR, followed by RFR, KNR, and the Linear model. This ranking aligns with the Pearson correlation coefficient and correlates well with MAE and MSE (cf. Table~\ref{TAB:ERROR} and Table~\ref{TAB:variogram}) confirming the hypothesis that NSR is a meaningful performance metric.

\section{Conclusion}

In this paper, we explored the potential and limitations of using end-to-end ML techniques for soil analysis with high-resolution air-coupled SFCW GPR measurements. Unlike previous studies~\cite{Castrignano2017} using sensor fusion and geostatistical methods to compare EMI and GPR and to delineate homogeneous zones in the field, our data-driven ML approach does not rely on domain-specific expert knowledge for calibration and allows for quantitative performance assessments. In this context, the nugget-to-sill ratio, which correlates with several standard ML performance metrics, has found to be a promising indicator for model performance, which does not require ground truth information. 

To establish SFCW GPR using ML-based processing as a geophysical tool for soil analysis suitable to precision farming, more application-specific datasets for supervised learning are needed. Multi-sensor field campaigns, such as the one reported here, are important to collect the large-scale labelled measurements needed for new instruments with manageable effort. The integration of additional sensors such as height sensors and
optical cameras could enable the identification of undesired
instrument sensitivities, including to driving conditions,
surface morphology and vegetation.  Incorporating remote sensing data~\cite{Gardin2021}, including UAV-based measurements~\cite{Bertalan2022}, could enhance data availability to cover a larger variety of soil types and conditions for the further development of methods.

\section*{Acknowledgment}

This work has been supported by Silicon Austria Labs (SAL), owned by the Republic of Austria, the Styrian Business Promotion Agency (SFG), the federal state of Carinthia, the Upper Austrian Research (UAR), and the Austrian Association for the Electric and Electronics Industry (FEEI).

\bibliographystyle{IEEEtran}
\bibliography{References}

\end{document}